\documentclass[english,aps,pra,twocolumn,showpacs,superscriptaddress,floatfix]{revtex4-1}
\usepackage[T1]{fontenc}
\usepackage[latin1]{inputenc}
\usepackage{amsmath}
\usepackage{graphicx}
\usepackage{amssymb}
\usepackage{dcolumn} 
\usepackage{bm} 
\usepackage[colorlinks=true,citecolor=blue,linkcolor=blue]{hyperref}
\usepackage{verbatim}
\makeatletter

\usepackage{amsfonts}
\usepackage{epsfig}
\usepackage{dsfont}
\usepackage{mathrsfs}
\usepackage{arydshln,leftidx,mathtools}

\newcommand{\Htil}{\tilde{\cal H}_{\rm aux}}
\newcommand{\Htilyz}{\tilde{\cal H}_{\rm aux}}
\newcommand{\Eq}[1]{Eq.~(\ref{#1})}

\newcommand{\be}{\begin{equation}}
\newcommand{\bea}{\begin{eqnarray}}
\newcommand{\eea}{\end{eqnarray}}
\newcommand{\ee}{\end{equation}}

\newcommand{\quot}[1]{``#1''}

\def\la{\langle}
\def\ra{\rangle}
\newcommand{\R}{{\mathcal R}}
\newcommand{\F}{{\cal F}}
\newcommand{\K}{\mathfrak K}

\usepackage{babel}
\makeatother

\begin{document}

\title{Adiabatic tracking of quantum many-body dynamics} 

\author{Hamed \surname {Saberi}}
\email{saberi@optics.upol.cz}
\affiliation{Department of Optics, Faculty of Science, Palack\'{y} University, 17. listopadu 12, 77146 Olomouc, Czech Republic}
\affiliation{Department of Physics and CeOPP, University of Paderborn, Warburger Stra{\ss}e 100, D-33098 Paderborn, Germany}

\author{Tom\'{a}\v{s} \surname {Opatrn\'y}}
\affiliation{Department of Optics, Faculty of Science, Palack\'{y} University, 17. listopadu 12, 77146 Olomouc, Czech Republic}

\author{Klaus \surname {M{\o}lmer}}
\affiliation{Department of Physics and Astronomy, University of Aarhus, DK-8000 Aarhus C,
Denmark}

\author{Adolfo \surname {del Campo}}
\affiliation{Department of Physics, University of Massachusetts Boston, Boston, MA 02125, USA}
\affiliation{Theoretical Division, Los Alamos National Laboratory, Los Alamos, New Mexico 87545, USA}
\affiliation{Center for Nonlinear Studies, Los Alamos National Laboratory, Los Alamos, New Mexico 87545, USA}

\date{December 5, 2014}

\begin{abstract}
The nonadiabatic dynamics of a many-body system driven through a quantum critical point can be controlled using counterdiabatic driving, where the formation of excitations is suppressed by assisting the dynamics with auxiliary multiple-body nonlocal interactions. We propose an alternative scheme which circumvents practical challenges to realize shortcuts to adiabaticity in mesoscopic systems by tailoring the functional form of the auxiliary counterdiabatic interactions. A driving scheme resorting in short-range few-body interactions is shown to generate an effectively adiabatic dynamics.
\end{abstract}

 	

\maketitle

{\it Introduction.} The adiabatic driving of quantum many-body systems is of interest to a wide variety of quantum technologies ranging from quantum simulation to adiabatic quantum computation. However, the implementation of adiabatic driving schemes in the laboratory is often challenging or simply impractical. In recent years a large body of theoretical and experimental progress has been focused on the development of shortcuts to adiabaticity (STA), fast-nonadiabatic protocols that reproduce the preparation of the same final state that would be achieved under slow driving~\cite{Torrontegui2013}. The feasibility of realizing such shortcuts relies on the control of nonadiabatic excitations away from the ground state manifold of the system of interest. A general technique to achieve this goal is known as counterdiabatic driving (CD)~\cite{Demirplak_Rice2003,Demirplak_Rice2005,Demirplak_Rice2008}. In a nutshell, the adiabatic approximation $|\Psi(t)\ra$ to the dynamics generated by a slowly-driven Hamiltonian of interest $H_0[\lambda(t)]$, is found to be the exact solution of the so-called counterdiabatic driving Hamiltonian $H_{\rm CD}$, i.e., $i\hbar\partial_t|\Psi(t)\ra=H_{\rm CD}|\Psi(t)\ra$, even under fast driving. More precisely, let the instantaneous eigenstates and eigenvalues of $H_0[\lambda(t)]$ be denoted by $\{|\varepsilon_n[\lambda(t)]\rangle\}$ and $\{\varepsilon_n[\lambda(t)]\}$, then the CD Hamiltonian can be always written as the sum $H_{\rm CD}=H_0[\lambda(t)]+H_{\mathrm{aux}}[\lambda(t)]$ of the system Hamiltonian $H_0[\lambda(t)]$ and the auxiliary CD term
\bea
\label{eq:H_aux}
H_{\mathrm{aux}}&=&i \hbar \dot{\lambda}(t) \sum_n[|\partial_\lambda \varepsilon_n\ra\la \varepsilon_n|-\la \varepsilon_n|\partial_\lambda \varepsilon_n\ra|\varepsilon_n\ra\la \varepsilon_n|] \; , \quad
\eea
where the over-dot denotes the time derivative. This auxiliary term suppresses explicitly excitations away from the adiabatic manifold allowing one to design  STA. Conversely, it vanishes as the adiabatic limit is approached~\cite{Demirplak_Rice2008,DelCampo_Rams_Zurek2012}. When applied to lattice systems~\cite{DelCampo_Rams_Zurek2012}, CD is closely related to the notion of quasiadiabatic continuation~\cite{HastingsWen05,Osborne07,VanAcoleyen13} and is equivalent to transitionless quantum driving~\cite{Berry2009}. 

To date, CD has been demonstrated in effective two-level systems~\cite{Bason2012,Ibanez2012,Zhang2013} but its general implementation is expected to be highly challenging in many-body systems. A remarkable experimental demonstration in a low-dimensional quantum fluid has recently been reported~\cite{Rohringer13}, but overall the technique appears to be restricted to self-similar processes~ \cite{Jarzynski13,DelCampoPRL13,DeffnerPRX14}. For few-particle systems additional progress has targeted the design of experimentally realizable variants of the CD Hamiltonian spanned by a set of implementable control fields~\cite{Opatrny2014}. 

Mimicking adiabatic dynamics is particularly challenging in many-body systems exhibiting a quantum critical point (QCP) $\lambda_c$ as a function of an external parameter $\lambda$. In the neighborhood of $\lambda_c$, the characteristic relaxation time and correlation length exhibit a power-law divergence as a function of the reduced parameter $\epsilon=[\lambda-\lambda_c]/\lambda_c$. As a result, the dynamics across QCP is expected to result in the formation of topological defects, signaling the breakdown of adiabatic dynamics~\cite{Dziarmaga10,PSSV11,DelCampo_Zurek14}. Signatures of critical dynamics are still present in finite systems and are masked by finite-size effects as the system size is reduced. Approaches to suppress defect formation in critical systems include nonlinear quenches~\cite{PB08,SSM08}, inhomogeneous driving~\cite{DR10,DR10b}, optimal control~\cite{Doria11,Caneva11} and multi-parameter tuning~\cite{SS14}, among other examples, see for a brief summary~\cite{DKZ13}. It has been shown that CD can be used to assist the adiabatic dynamics across a quantum phase transition at the cost of engineering $H_{\mathrm{aux}}$ which involves multiple-body non-local counterdiabatic interactions~\cite{DelCampo_Rams_Zurek2012}. As a result, implementing $H_{\mathrm{aux}}$ in the laboratory remains challenging despite recent progress in digital quantum simulation~ \cite{Muller11,Barreiro11}.

In this Rapid Communication, we propose a practical recipe for suppression on demand of the nonadiabatic transitions in arbitrary many-particle systems. We illustrate the applicability of our method in the context of spin chains. Our scheme is of particular relevance to state preparation in digital quantum simulators, e.g. in trapped ion chains~\cite{Muller11,Barreiro11,Casanova12} and Rydberg gases~\cite{Weimer10}.

{\it Counterdiabatic driving of many-body systems.} We consider a quench of a \emph{finite} transverse Ising chain of $N$ spins with nearest-neighbor interactions under open boundary condition described by 
\begin{eqnarray}
\label{eq:trans_Ising}
H_0(t) = -B(t) \sum_{j=1}^N \sigma_{j}^x+J_0 \sum_{j=1}^{N-1} \sigma_j^z\otimes\sigma_{j+1}^z \; ,
\end{eqnarray}
where $B(t)$ is a time-varying external magnetic field, $\sigma_j^x, \sigma_j^z$ are usual Pauli sigma operators associated with site $j$, and $J_0$ denotes uniform magnetic couplings between adjacent spins assumed to be unity throughout the work. In the thermodynamic limit the system exhibits a quantum phase transition as a function of the transverse field, separating a paramagnetic phase ($|B(t)/J_0|>1$) 
from a doubly-degenerate antiferromagnetic one ($|B(t)/J_0|<1$).   

\begin{figure}[b]
\centering
\includegraphics[width=0.8\linewidth]{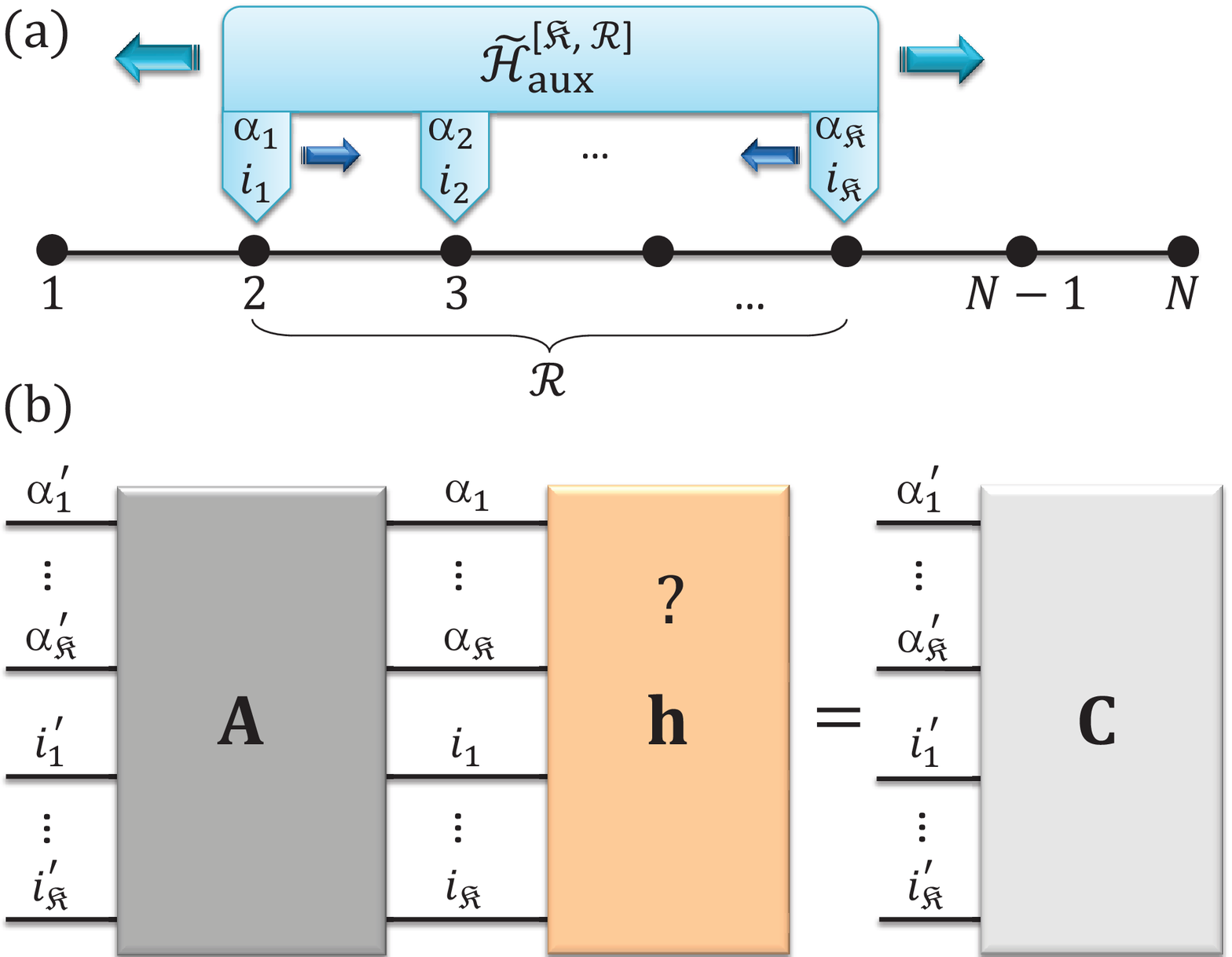}
\caption{(Color online) (a) Schematic of the variational construction of the ansatz of the form \Eq{eq:Htil} to the \quot{exact} auxiliary CD term in \Eq{eq:H_aux} for possible experimental implementation in trapped ion chain. $\Htil^{[\K,\R]}$ corresponds to the action of an itinerant \quot{train} of $\K$-partite Pauli interactions between all possible permutations of $\K$-tuple of spins within the range $\R$ and with optimal amplitudes $h_{i_1,\ldots,i_{\K}}^{\alpha_1,\ldots,\alpha_{\K}}$ obtained from the contraction pattern depicted in (b).
}
\label{fig1_contract}
\end{figure}

The quantum simulation of this model and its variants is the goal of current efforts in ion-trap experiments~ \cite{Friedenauer08,Kim10,Kim11,Richerme13}, ultracold atoms~\cite{Simon11}, and NMR experiments~\cite{Li14}, to name just a few examples. The achievable system-size with current technology is still tractable by real-space exact diagonalization (ED), which we employ to derive the instantaneous eigenspectrum $\{|\varepsilon_n(t)\rangle\}$ and the auxiliary CD term via \Eq{eq:H_aux}. $H_{\mathrm{aux}}$ is found to involve $\K$-body interactions ($\K=1,\dots,N$), in agreement with~\cite{DelCampo_Rams_Zurek2012}. We wish to find an alternative CD scheme resorting exclusively in auxiliary control fields associated with a \emph{restricted} set of operators, assumed to be available in a quantum simulator. The scheme relies on suggesting a new ansatz for the variational construction of the \quot{exact} auxiliary CD term of the form
\bea
\label{eq:Htil}
\Htil^{[\K,\R]}(t)=\sum_{i_1,\ldots,i_{\K}}^{'}\sum_{\alpha_1,\ldots,\alpha_{\K}} h_{i_1,\ldots,i_{\K}}^{\alpha_1,\ldots,\alpha_{\K}}(t)\;\;
\bigotimes_{\ell=1}^{\K} \sigma_{i_{\ell}}^{\alpha_{\ell}} \; ,
\eea
where $0 < \vert i_{\ell}-i_{{\ell}'}\vert \le \R$ $\forall {\ell},{\ell}'$ is understood in the restricted sum over distinct site indices at maximal distance $\R=\{\K-1,\dots,N-1\}$ from each other and to be identified as the \emph{range} of the $\K$-body interaction, $\alpha_{\ell}=\{0,x,y,z\}$ with $\sigma^0\equiv\mathds{1}$, and $h_{i_1,\ldots,i_{\K}}^{\alpha_1,\ldots,\alpha_{\K}}$ are the optimal interaction amplitudes to be found via a variational optimization procedure that shall be detailed in the following. This ansatz is general enough to generate an arbitrary unitary. However, we shall see that a $\Htil^{[\K,\R]}$ containing exclusively few-body quasi-local terms suffices to induce an effectively adiabatic dynamics. $\Htil^{[\K,\R]}$ involves \emph{nonlocal} $\K$-body couplings of spins through the chain as Fig.~\ref{fig1_contract}(a) illustrates. Considering the driving of an arbitrary eigenstate $|\Psi(0)\ra=|\varepsilon_n(0)\ra$ prepared at $t=0$, the optimal amplitudes $h_{i_1,\ldots,i_{\K}}^{\alpha_1,\ldots,\alpha_{\K}}(t)$ that lead to an optimal representation of $H_{\rm aux}$ within the subspace of the ansatz (\ref{eq:Htil}) may be obtained by minimizing the quadratic cost function of the form
\begin{eqnarray}
\label{eq:cost}
\min_{h_{i_1,...,i_{\K}}^{\alpha_1,...,\alpha_{\K}}(t)} \| (H_{\mathrm{aux}} - \Htil^{[{\K},\R]}) |\Psi_{\rm a}(t)\rangle \|^2 \; , 
\end{eqnarray}
where $|\Psi_{\rm a}(t)\ra \equiv |\varepsilon_n(t)\ra$ denotes the adiabatic evolution of $|\Psi(0)\ra$ generated by $H_0(t)$. The result of such a minimization can generally be written down in the form
\begin{eqnarray}
\label{eq:lin_algeb_eqs}
\sum_{\{i_{\ell}\}}^{'} \sum_{\{\alpha_{\ell}\}} A_{i_1',\dots,i_{\K}', i_1,\dots, i_{\K}}^{\alpha_1',\dots,\alpha_{\K}', \alpha_1,\dots,\alpha_{\K}}   
h_{i_1,\dots,i_{\K}}^{\alpha_1,\dots,\alpha_{\K}}= C_{i_1',\dots,i'_{\K}}^{\alpha_1',\dots,\alpha'_{\K}}\; ,
\end{eqnarray}
where
\begin{eqnarray}
\label{eq:A_C}
\nonumber
 & & A_{i_1',\dots,i_{\K}', i_1,\dots, i_{\K}}^{\alpha_1',\dots,\alpha'_{\K}, \alpha_1,\dots,\alpha_{\K}}\!\! =\!  
{\rm Tr}\!\left[\rho_t^{\rm a}  \big\{\bigotimes_{{\ell}=1}^{{\K}} \sigma_{i_{\ell}}^{\alpha_{\ell}},\bigotimes_{{\ell}=1}^{{\K}} \sigma_{i'_{\ell}}^{\alpha'_{\ell}} \big\}\!\right],
 \\
& & C_{i_1,\dots,i_{\K}}^{\alpha_1,\dots,\alpha_{\K}}= {\rm Tr}\left[\rho_t^{\rm a}  \big\{H_{\mathrm{aux}}, \bigotimes_{{\ell}=1}^{\K} \sigma_{i_{\ell}}^{\alpha_{\ell}}\big\} \right] \; ,
\end{eqnarray}
where the anticommutator $\{A,B\}=AB+BA$ and we have denoted $\rho_t^{\rm a}=|\Psi_{\rm a}(t)\rangle \langle \Psi_{\rm a}(t)|$ to stress the flexibility of the formalism in extending to mixed states. At a given time $t$, the equations above admit a compact form as a single linear tensor equation involving a \emph{contraction} of the form 
\begin{eqnarray}
\label{eq:contraction}
\overbracket{\mathbf{A}\mathbf{h}} = \mathbf{C} \; ,
\end{eqnarray}
with a graphical representation depicted in Fig.~\ref{fig1_contract}(b). 

The latter tensor equation can be solved by an in situ reshaping of the tensor $\mathbf{A}$ of rank $4{\K}$ into a regular matrix $\bar A$ of dimension $(4N)^{\K}\times (4N)^{\K}$ while combining the set of indices $\{i_{\ell}\}\{\alpha_{\ell}\}$ and $\{i'_{\ell}\}\{\alpha'_{\ell}\}$ into single \quot{superindices} $I$ and $I'$, respectively. This amounts to casting the original tensor equation into a matrix inversion problem involving matrix ${\bar{A}}_{I',I}$ and reshaped column vectors ${\bar{h}}_I$ and ${\bar{C}}_{I'}$ given by $\bar h = {\bar{A}}^{-1} \bar{C}\; $. Reshaping back properly the column vector $\bar h$ so obtained into the original order yields the optimal amplitudes $h_{i_1,...,i_{\K}}^{\alpha_1,...,\alpha_{\K}}$ for approximating the exact auxiliary CD term. A costly part of the implementation of the outlined procedure requires an explicit construction of $\bar{A}$ and $\bar C$ which in case of the driving of pure states can be economized on by utilizing the following properties and bringing thereby a significant reduction in CPU time: (i) $\bar{A}_{I',I}=\bar{A}_{I,I'}$; (ii) simplifying the definitions in \Eq{eq:A_C} as
\begin{subequations}
\label{eq:A_C_rewritten}
\begin{eqnarray}
\hspace{-20mm} {\bar{A}}_{I',I}  &=& 2 \mathrm{Re}\{\la \Phi_{I'}|\Phi_I\ra\}\; , \\
\bar{C}_{I} &=& 2 \mathrm{Re}\{\la \Phi_I| \Phi_{\rm aux}\ra\} \; ,
\end{eqnarray}
\end{subequations}
where $|\Phi_I\rangle \equiv \bigl(\bigotimes_{\ell=1}^{\K} \sigma_{i_{\ell}}^{\alpha_{\ell}}\bigr) |\Psi_{\rm a}(t)\rangle$ and $|\Phi_{\rm aux}\rangle \equiv H_{\rm aux} |\Psi_{\rm a}(t)\rangle$.

\begin{figure}[!h]
\includegraphics[width=0.89\linewidth]{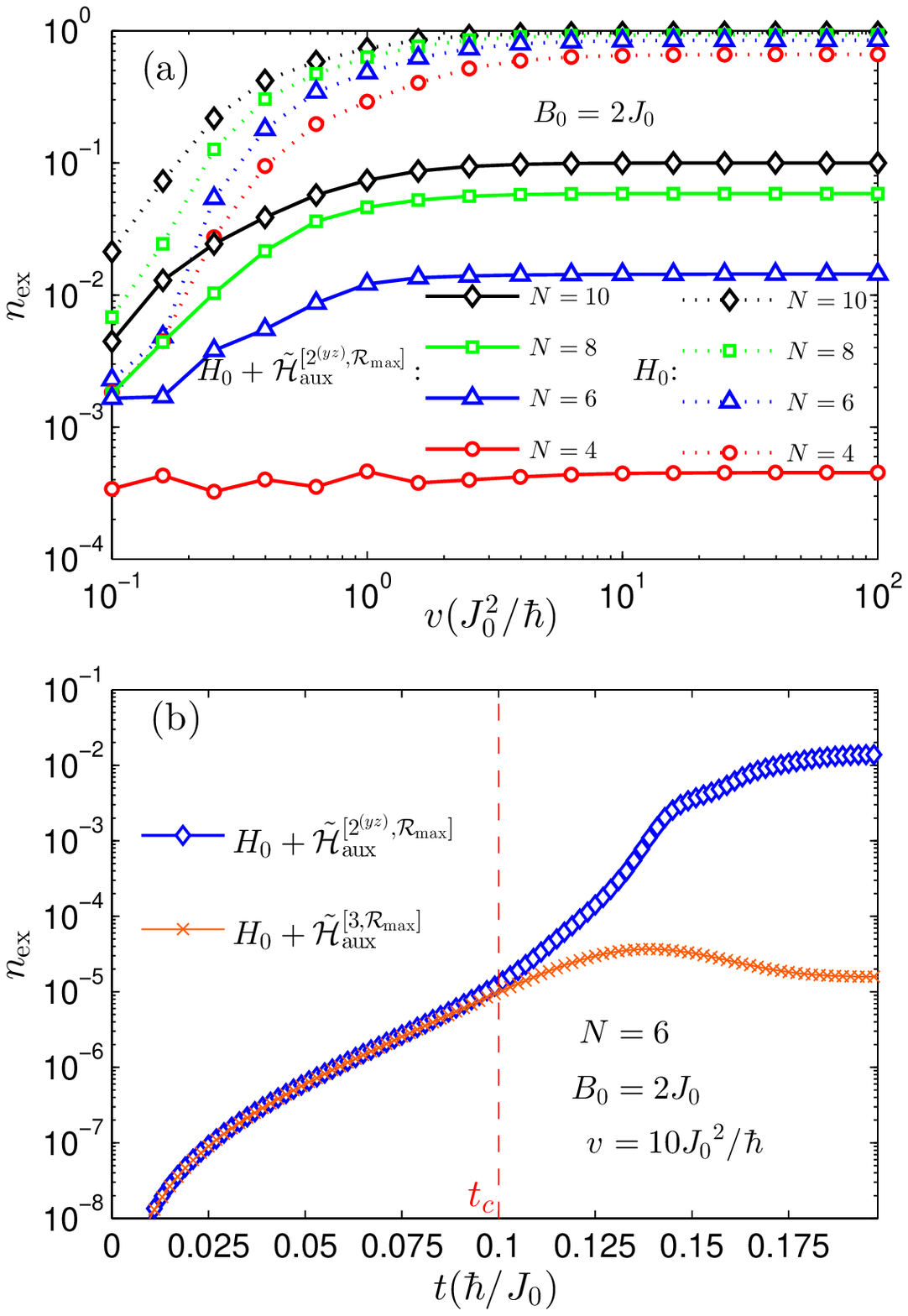}
\caption{(Color online) (a) Suppression of DoE following the passage through the QCP induced by the variational ansatz restricted to two-body interactions, as in (\ref{eq:H_C_two_body}), as a function of the quench rate $v$ in a linear quantum Ising chain for different system sizes. At fast quench rates, the efficiency of the ansatz in suppressing excitations saturates and a further suppression can be achieved by including higher-order multiple-body terms, as (b) depicts. 
}
\label{fig2_KZM}
\end{figure}

The ansatz (\ref{eq:Htil}) is fairly general and its implementation in a quantum simulator can be expected to be complex. In what follows we show that short-range few-body interactions suffice to generate an effectively adiabatic dynamics. For the sake of illustration, we start discussing the case in which $\Htil^{[\K,\R]}$ is 
restricted to two-body interactions and show that it already suffices to reduce the density of excitations (DoE) by orders of magnitude with respect to the $H_{\rm CD}=H_0$ case. After discussing it, we shall include higher-order multiple-body interactions.

\begin{figure*}[t]
\centering
\includegraphics[width=0.91\linewidth]{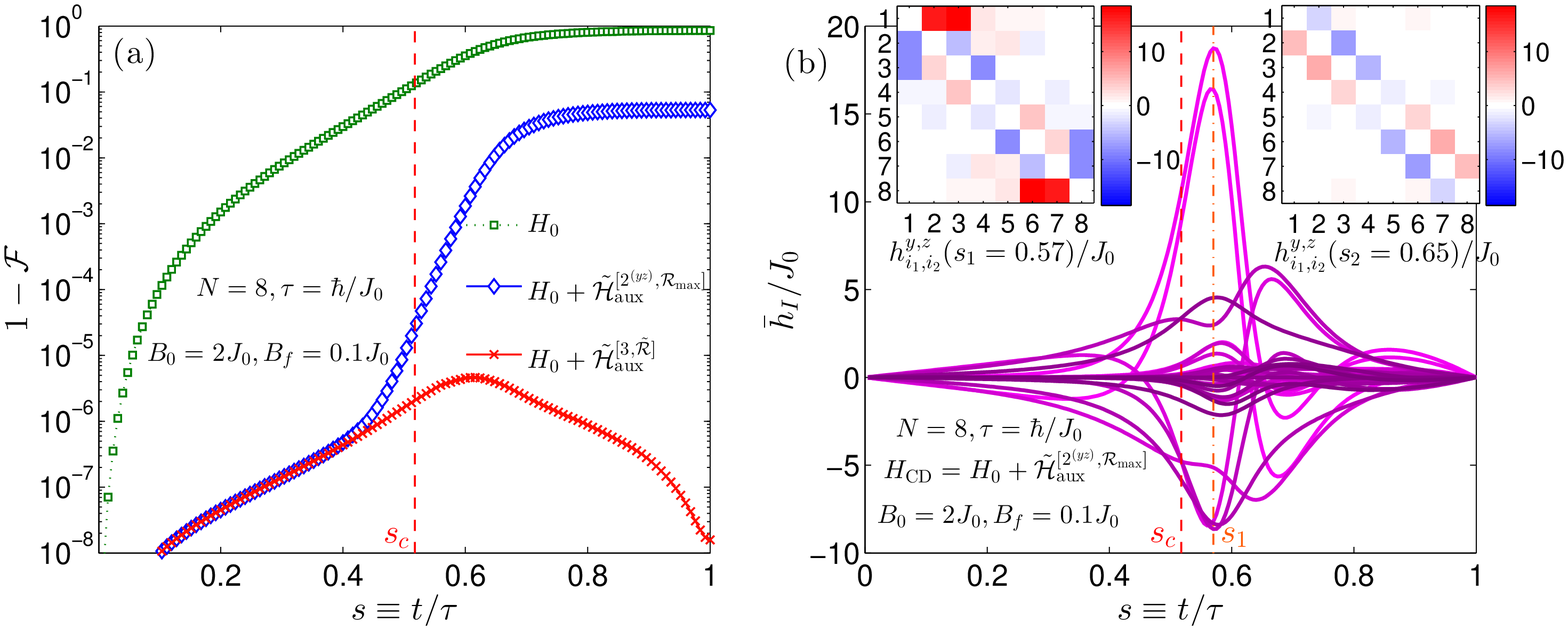}
\caption{(Color online) (a) Time-evolution of the state preparation infidelity during a shortcut to the adiabatic driving of the ground state of a quantum Ising chain. The fidelity can be improved significantly by incorporating higher-body interactions at a truncated range of $\tilde{\R}=\R_{\max}-4$. (b) Real-time flow of the interaction amplitudes associated with the implementation of the full-range two-body ansatz as in \Eq{eq:H_C_two_body} during passage through QCP of the model denoted by $s_c$. The color maps in the insets visualize the strength of two-body interactions $h_{i_1,i_2}^{y,z}$ among all pairs of spins at sites $(i_1,i_2)$ in the chain.
}
\label{fig3_state_prep}
\end{figure*}

As an illustrative case, and without loss of generality, we shall henceforth focus on the driving of the ground state of the initial system Hamiltonian, $|\Psi(0)\ra=|\varepsilon_0(0)\ra$. The quality of the adiabatic tracking scheme can be estimated by means of the time-dependent fidelity
\begin{eqnarray}
\label{eq:CD_fidelity}
\F(t) \equiv |\langle \Psi_{\rm a}(t)|\tilde{\Psi}(t)\rangle|^2 \; ,
\end{eqnarray}
defined as the overlap between the adiabatic evolution of the starting state and the instantaneous state of the system denoted by $|\tilde{\Psi}(t)\ra$ under an evolution dictated by $H_0(t)+\Htil^{[\K,\R]}(t)$ while using the units in which $\hbar=1$. The DoE $n_{\mathrm{ex}}$ can then be computed as the probability of not ending up in the final ground state, i.e., $n_{\mathrm{ex}}\equiv1-\F$. We consider a quench linear in time of the form $B(t)=B_0-vt$ and study the efficiency of $\Htil^{[\K,\R]}$ in suppressing excitations as a function of the quench rate $v$ for different system sizes following the paradigm of Kibble-Zurek mechanism~\cite{Dziarmaga10}. 

Figure~\ref{fig2_KZM} summarizes the results with the simplest possible choice of a subclass of the proposed variational ansatz in \Eq{eq:Htil} of the form
\begin{eqnarray}
\label{eq:H_C_two_body}
{\Htilyz}^{[2^{(yz)},\R_{\rm max}]}(t) \equiv \sum_{i_1,i_2}^{'} h_{i_1, i_2}^{y,z}(t) \; \sigma_{i_1}^y \otimes\sigma_{i_2}^z \; ,
\end{eqnarray} 
motivated by our observation that for $\K=2$ other permutation of spin components do not contribute to the suppression of excitations. We take the $H_{\rm CD}=H_0$ case as reference, where $n_{\mathrm{ex}}$ increases monotonically until reaching saturation due to finite-size effects~\cite{Dziarmaga05,ZDZ05,Polkovnikov05}. It is shown that a variational ansatz restricted to only two-body terms though with an interaction range extending through the whole chain, i.e., $\R_{\max}=N-1$, leads to a successful suppression of $n_{\mathrm{ex}}$ by several orders of magnitude depending on the system size. Here the system is evolved in time from an initial value of the quench parameter $B_0<B(t_c)$ to a final one deep in the ferromagnetic phase where the value of $n_{\rm ex}$ in the plot is collected. The results therefore suggest a high-fidelity adiabatic passage across the QCP of the model at $t_c=(B_0-1)/v$. The residual excitations can be further suppressed by higher-order multiple-body interactions, as shown in Fig.~\ref{fig2_KZM}(b). 

{\it Shortcuts to adiabatic state preparation.} We demonstrate that the variational counterdiabatic ansatz can be used to achieve high-fidelity state preparation. We consider for this purpose a time-dependent modulation of the magnetic field which enforces $H_{\rm aux}$ to vanish at the beginning and end of the driving scheme, $t=0,\tau$. We are thus led to the boundary conditions $B(0)=B_0$, $B(\tau)=B_f$ and $\dot{B}(0)=0$, $\dot{B}(\tau)=0$ which are satisfied by a polynomial quench of the form 
$B(s)=B_0+ 3(B_f-B_0)s^2 - 2(B_f-B_0)s^3$ with $s \equiv t/\tau$. The use of the latter quench is further motivated by adiabatic perturbation theory~\cite{Lidar09}. Figure~\ref{fig3_state_prep} illustrates the results for preparation of the ground state of the transverse Ising chain in \Eq{eq:trans_Ising} under such a quench and for various choices of $\Htil^{[\K,\R]}$. We use the fidelity in \Eq{eq:CD_fidelity} to assess the quality of the preparation procedure. With a full-range two-body interaction, the protocol leaves some room for improvement, as shown in Fig.~\ref{fig3_state_prep}(a), indicating the urge to employ higher-body terms for longer chain lengths. However, for higher-body interactions beyond $\K=2$ the computational complexity associated with the large number of spins permutations and thereby the dimensions of the matrices to be formed and inverted renders the numerical implementation of the variational procedure intractable. To circumvent such a practical challenge, we suggest to truncate over the range of the $\K$-body ansatz by restricting it to manageable values of $\tilde{\R} \ll \R_{\max}$. The latter truncation strategy is further motivated by the nearly tridiagonal structure of $h_{i_1,i_2}^{y,z}$ evident in the inset of Fig.~\ref{fig3_state_prep}(b) which shows remarkably that the dominant contributions to $\Htil^{[2^{(yz)},\R_{\max}]}$ consist of only short-range interactions. Figure~\ref{fig3_state_prep}(a) illustrates the success of the truncation strategy by demonstrating significant improvement in fidelity upon employing short-range three-body interactions. We point out the number of distinct $\K$-body interactions as the required experimental resources to achieve the maximal-fidelity state preparation within our scheme scales with ${1\over2}{4^{\K}N! \over (N-{\K})!}$ which is a polynomial in the size of the system $N$ of the leading order $\mathcal{O}(N^{\K})$. The scaling  derives from a simple combinatorics corresponding to the total number of possible choices of a $\K$-tuple of spins from $N$ ones in which internal permutation of tuples produces distinct choices due to noncommutativity of the Kronecker product. The prefactors $1\over2$ and $4^{\K}$, moreover, account for the mirror symmetry of the finite open chain under consideration and the multiplicity associated with various spin components of a Pauli sigma operator, i.e., $\{0,x,y,z\}$, respectively.

{\it Conclusions.} We have shown how to engineer an experimentally realizable counterdiabatic control Hamiltonian for the fast driving of many-body spin systems that mimic adiabatic driving. Our approach combines ED with a variational principle to determine the optimal CD scheme with a restricted set of control fields and leads to a suppression of the DoE by several orders of magnitude with respect to the uncontrolled driving dynamics. Although the identification of $H_{\rm{aux}}$ is a computationally hard problem that poses a challenge to scalability of the method, tests on finite systems are still relevant to currently feasible experiments and the variational approximations to the CD Hamiltonians for these cases may guide future approaches towards large particle numbers. In congruence with recent results in optimal control theory~\cite{Kallush2011,Lucas2013,Lloyd2014,Wu2014}, our results suggest that the practical implementation of our scheme represents an effort scaling only \emph{polynomially} with the system size. 
Our proposal is ideally suited for digital quantum simulation as well as tailoring the nonequilibrium thermodynamics of many-body systems~\cite{Dorner12}. It further supplements previous adiabatic tracking schemes aimed at accessing highly excited states~\cite{MolinerPRL13}. The possibility to reduce the level of nonlocality in implementing the CD term promised by our approach may facilitate realization of adiabatic quantum computation~\cite{Rezakhani2009,Rezakhani2010}. 

We acknowledge stimulating discussions with Daniel Lidar, Marek M. Rams, Alexey Gorshkov, and Xi-Wen Guan.
H.S. is grateful to Aarhus University for support and hospitality. This work was financed by the European Social Fund and the state budget of the Czech Republic, project CZ.1.07/2.3.00/30.0041. This research is further supported by the U.S. Department of Energy through the LANL/LDRD Program and a LANL J. Robert Oppenheimer fellowship (AD).

\bibliography{MB_adiabatic}

\end{document}